\documentclass[12pt]{article}
\usepackage[dvips]{graphicx}

\begin{document}
\title{Effective interaction of exhausted regions in
kinetics of nucleation
}
\author{Victor Kurasov}
\date{  }

\maketitle

\begin{abstract}
An effect of overlapping of exhausted regions around
droplets of a new phase is analyzed. Several approximation
to take overlapping into account are suggested. The small
parameter responsible for convergence is extracted. The
multi particle effects of overlapping are described.
\end{abstract}

\section*{Introduction}

Kinetics of the first order phase transitions includes
many aspects of evolution. One of such aspects is the
exhaustion of the volume where the embryos can appear.
Since A.N.Kolmogorov \cite{Kolm} has solved a problem
of exhaustion for the case of crystallization this problem
is considered to be analyzed completely.

Nevertheless in the case of nucleation of a supersaturated
vapor into a liquid state of droplets
kinetics of nucleation is governed by another mechanisms
which depends on the Knudsen number characterizing the
nucleating system. Under the small Knudsen numbers
the kinetics of nucleation was described in \cite{Kuni}.
Under the big Knudsen numbers kinetics resembles kinetics
of crystallization but with some modifications. These
modifications were described in \cite{PhysicaA}. Here the
precise solution of this problem is absent. The source of
difficulties lies in a very specific behavior of exhausted
regions. In this paper the further investigation of this
problem is presented.

\section{Direct regions}

The rate of nucleation  $J$ as it is shown in \cite{PhysicaA} is
proportional to the volume $V_{free}$ of the non-exhausted
 region
 $$
 J \sim V_{free}
 $$
 The total volume of the system $V_{tot}$ is supposed to be
 the unit one. Our task is to determine $V_{free}$ as
 accurate as possible.

Certainly,
$$
V_{free} = 1 - V_{exh}
$$
where
$V_{exh}$ is the volume of the region exhausted by already
existed droplets.

Around the formed droplet there will be an exhausted region
formed by the vapor consumption by this very droplet.
This region will be called as a "direct region".

For $V_{exh}$ one can write the following representation:
$$
V_{exh} = \int_0^t J(t') V_{cl}(t,t') dt'
$$
Here $t=0$ is the initial moment of time, $t$ is the
current moment of time. The value $V_{cl}(t,t')$  is the (mean)
volume which is exhausted at the moment $t$
from $V_{free}$ by the embryos
formed at $t'$.
For the last value one can suggest several approximations.

\subsection{Kolmogorov's result}

In crystallization Kolmogorov justified the following
result
$$
V_{cl} (t, t') =  V_0 (t, t')
$$
 Here $V_0(t,t')$ is the volume of the region
which is exhausted
at $t$
elsewhere by embryos formed at $t'$.

In crystallization
$$
V_0 \sim (t-t')^d
$$
where
$d$ is the dimensionality of space.
The linear size $R \sim V_o^{1/d}$
of the
direct region grows linearly in time. This is the crucial
point in justification  of the Kolmogorov's result.

In general situation of nucleation
asymptotically at $t-t' \rightarrow \infty$ one can see
that
$$
O(t-t')^{d/2} \leq V_0 \leq O(t-t')^d
$$

It means that the linear size  grows as
$$
O(t-t')^{1/2} \leq R \leq O(t-t')
$$
So, the non-linearity is not too strong and
this will be the source of several approximations considered
below.

In any situation the dependence $R(t-t')$ is known.
Ordinary $R \sim (t-t')^{1/2}$.

\subsection{Stochastic approximations}

In stochastic approximation  the following approximation
$$
V_{cl} (t, t') = V_{free}(t) V_0 (t, t')
$$
is taken.

This approximation implies stochastic overlapping of
volumes $V_0$ coming from different droplets.
Why for the volume $V$ depending oh $t$ and $t'$ the
overlapping is taken at $t$? More reasonable is to take the
half-sum
$$
V_{cl} (t, t') = \frac{(V_{free}(t) + V_{free}(t'))}{2}
 V_0 (t, t')
$$
or
$$
V_{cl} (t, t') = \int_{t'}^t V_{free}(t'') dt''
 V_0 (t, t')
$$

The approximate linearity of $R(t-t')$ leads to the
approximate similarity of all approaches.

\section{Approximation of pair interaction}

Consider the auxiliary problem which is formulated as
following:

{\it
Consider the droplet born at $t''$. The question is what
will be the part of $V_0$ occupied by direct regions from other
droplets.
}

One can approximately suppose that outside direct region of
this droplet the rate of nucleation is not perturbed and
inside the direct region the rate of nucleation is zero.

Then the problem is purely geometrical one. One has simply
to calculate the overlapping of spheres.
The value $V_{over}(r'',t'',r',t')$ is the volume of
overlapping for the droplet born at $t''$ in the point
$\vec{r''}$ and the droplet born at $t'$ in the point
$\vec{r'}$. It is calculated purely geometrically.

Also purely geometrically one can calculate the boundary
$r_b'$ of the direct region and the time $t_b$ when the
given $r''$ is attained by the boundary of this direct
region.

As the result one can easily calculate the total volume of
overlapping by integration
$$
V_{ol} = \int_0^{t_b} dt' \int_0^{2R(t-0)} dr S_d
V_{over}(0,t'',r.t')
$$
Here $t_b$ is a minimum of $t$ and $t_b$,
$S_d$ is the surface square of a sphere of radius $r$ in
a space with dimension $d$.

The problem is that the supposition that
 outside direct region of
the droplet the rate of nucleation is not perturbed and
inside the direct region the rate of nucleation is zero is
no more than an approximation.

Another approximation is that the rate of nucleation
outside is $J(t')$ taken from another approximation, for
example, from the previous  one. Then
$$
V_{ol} = \int_0^{t_b} dt' \int_0^{2R(t-0)} dr S_d J(t')
V_{over}(0,t'',r.t')
$$

Unfortunately, the rate of nucleation $J(t')$ strongly depends
here on $r$. To see this we can analyze the region (it will
be called "an inverse region") where the region is
forbidden to have the free point $r,t'$. This region can
cross the direct region for the droplet born at $0,t''$.

It means that in the outside region near $r_b'$ the rate of
nucleation is higher than far from $r_b'$

This correlation is missed here.

\section{Approximations for estimates}

The effective approach is to note that $d \gg 1$ and use
this fact. For example for $d=3$ the droplets appeared
until the first quarter of the nucleation period in decay
conditions occupy the small part of the whole volume and
later all evolution is governed namely by these droplets.
This fact allows to construct approximations as it is shown
in \cite{book}.

\section{Inverse region}

Inverse region for the droplet born at $r',t'$ is the
geometrical place of points where the formation is
forbidden to have the possibility to form droplet at
$r',t'$.
Certainly, this region is situated symmetrically to the
direct region of the same droplet.

In the case of the Kolmogorov's solution $R = A (t-t')$ with
some constant $A$. Then any inverse region can not cross
any direct region.

Having required that at the bottom of inverse region there
is no formation of droplets one can see that there is no
traces of direct regions in the whole inverse region. Then
the Kolmogorov's approximation really takes place.

Consider now the general situation. Let $P_-(t)$ be a
probability that in the point $r=0$ at $t$ there is a free
point.
For $P_-(t)$ one can write
$$
P_-(t) = \int_0^t dt' \int_0^{r_b} dr' S_d P_-(t',r')
$$
where
$P_-(t',r')$ is the probability that the point $r',t'$ is
free.

Let $r',t'$ be inside th inverse region formed at $r=0,t$.
For $P_-(t',r')$ one has to construct the inverse region.
This region can go outside the inverse region of $r=0,t$.
This is the  difficulty.
For $P_-(t',r')$ one
can then write
$$
 P_-(t',r') =
\int_0^{t'} dt'' [ \int_{V_{exhausted}} d \vec{r''}
P_-(t'',r'') +
\int_{V_{non-exhausted}} d \vec{r''}
P_-(t'',r'') ]
$$
Here $V_{exhausted}$ is the inverse region based on $r',t'$
which lies inside the inverse region based on $0,t$ and
$V_{non-exhausted}$ is the inverse region based on $r',t'$
which lies outside the inverse region based on $0,t$.

The last separation is artificial but it will be essential
in further approximations.

The integral equation is the linear one. Moreover, in
dimension $d=3$ there exists a cylinder symmetry.

The suitable way is an iteration solution based on the
smallness of overlapping
$$
 P_{-(i)}(t',r') =
\int_0^{t'} dt'' [ \int_{V_{exhausted}} d \vec{r''}
P_{-(i-1)}(t'',r'') +
\int_{V_{non-exhausted}} d \vec{r''}
P_{-(i-1)}(t'',r'') ]
$$
 The initial approximation is
evident: $P=1$ in $V_{exhausted}$,  $P=P_{mean}$ in
$V_{non-exhausted}$. Here $P_{mean}$ is some mean rate of
nucleation calculated by other approximations.
The
smallness of overlapping
ensures the convergence of iterations.

This consideration is also some approximation. Here we do
not take into account the possibility to born the third
droplet with its own inverse region which causes the change
of $P_-$.

\section{Approximate simplification}

If one approximates $J$ by initial intensity $J_0$ in
$V_{exhausted}$ and by the mean intensity $\bar{J}$ in
$V_{non-exhausted}$
then one can see a very simple expression
$$
J(t',r') =
\int_0^{t'} dt'' [J_0 V_{exhausted}
+\bar{J}V_{non-exhausted} ]
$$
The last approximation is both simple and accurate one.

Earlier the effects of interaction (overlapping) of three
and more regions were not taken into account.
Since the effects of overlapping are small, the influence
of effective interaction of several particles are small.

\section{One limit case}

In order to construct the procedure taking into account
effects of overlapping (interference) of exhausted regions
initiated by different droplets one has to outline the
parameter responsible for the relative smallness of the
many-droplets effects. Only then the effects under
consideration will have the order of corrections and can be
considered as some perturbations.

The small parameter $\epsilon$ of the theory will be, thus,
a ratio of the square of exhausted region with
overlapping to the the square of exhausted region without
overlapping. We expect that $\epsilon$ is seriously less than $1$.

Ordinary the growth of the linear size of the exhausted
region can be well approximated by a power-like dependence
$$
R(t,t') = A (t-t')^{\alpha}
$$
with parameters $A$ and $\alpha$. The case $\alpha
\rightarrow \infty$ corresponds to the maximum of
$\epsilon$.

Fortunately, this case allows analytical solution.

The case $\alpha = \infty$ corresponds to instantaneous
creation of the exhausted region of some radius $R_0$.
Then the probability $P_-(r,t)$ to be the point $r$ free at
the moment $t$ equals to
$$
P_- (r,t) = \int_{S_{R_0}} dr' \int_0^t dt'  P_- (r',t') I_0
$$
where the space integral is taken over the sphere
of the radius $R_0$ with the
center $r$ and $I_0$ is the kinetic factor, i.e. the
free intensity of the droplets formation.

The crucial point of consideration is that here $P_-$ can
not depend on $r'$. Then
$$
P_- (r,t) =  4 \pi R_0^3 \int_0^t dt'  P_- (r,t') I_0 /3
$$

The differentiation gives
$$
\frac {\partial P_- (r,t) }{\partial t }
=  4 \pi R_0^3   P_- (r,t) I_0 /3
$$

The last differential equation has an evident solution
$$
P_- (r,t) = exp(-4 \pi R_0^3 t / 3)
$$
where initial moment of time is taken as $t=0$.

Then it is easy to calculate the total number of exhausted
regions
$$
M_{tot} = \int_0^\infty dt P_- (r,t)
=\int_0^\infty dt
exp(-4 \pi R_0^3 t / 3) = 3/(4 \pi R_0^3)
$$
The absence of overlapping corresponds to the linearized
version, i.e.
$$
M_{lin} =
=\int_0^{3/(4 \pi R_0^3)} dt
(1-4 \pi R_0^3 t / 3) = 3/(8 \pi R_0^3)
$$
So, $M_{tot} = 2 M_{lin}$. Since the squares of all
exhausted regions are $4 \pi R_0^3  / 3$, the last ratio
will be the power of overlapping, i.e.
$$
\epsilon = (M_{tot} - M_{lin})/M_{lin}
$$

So, in any situation $\epsilon$ and can be considered as
a small parameter.

\section{The scheme of calculation}

Now it is worth to present a concrete effective scheme of
construction of the multi-particle effects of overlapping.

The first step is the formal one - to form the first
droplet at $r', t'$.

The rate of nucleation inside the exhausted region of the
first droplet ($ER_1$) is zero. The question is to find the
rate of nucleation outside $ER_1$.

Then it is possible to choose two alternative possibilities
to go forward. The first one is to choose the rate
of nucleation outside $ER_1$ to be the ideal value $I_0$.
This approach is  very close to the direct simulation of
the process. Unfortunately, in this approach to get a
suitable effect it is necessary to take into account many
neighbor droplets. For example, in the plane case the number
of neighbors has to be greater than six. In the
free-dimensional case this number is greater than twelve.

The second possibility is more effective. Instead of $I_0$
outside the $ER_1$ one can use the average rate of
nucleation $\bar{I}$. One can define the average rate of
nucleation $I_V $ over the  volume $V$ as
$$
I_V  = \frac{1}{V} \int_V dr I(r,t)
$$

The value $\bar{I}$ is the limit value
$$
\bar{I} = I_V
$$
Namely this value  is chosen as the rate of nucleation
outside $ER_1$.

Corrections to the last value appear because near the
boundary of $ER_1$ some droplets which might affect on a
point outside $ER_1$ had been born in $ER_1$ where formation is
forbidden.

The rate of this value has an order of relative volume of
intersection of $ER_1$ and inverse region $IR$ of the
second droplet, i.e. $IR_2$. More precise the probability
$P_2$ to form the droplet at $r'',t''$ when the first
droplet is formed at $r',t'$
$$
P_2 (r'',t'';r',t') =
\frac{\bar{I}}{I_0} [ 1 - \frac{ Vol(ER_1 \cap
IR_2) }{Vol( IR_2)} ]
$$
Here $Vol$ means the volume. The value $  Vol(ER_1 \cap
IR_2)$ is the volume of the intersection
of $ER_1$ and $IR_2$, etc.

The volumes standing here are the simple known algebraic
functions, the calculation is evident.

The effectiveness of this procedure is based on the relative
smallness of the value
$$
\sigma =
\frac{ Vol(ER_1 \cap
IR_2) }{Vol( IR_2)}
$$

The average probabilities are given by
$$
P_2(r'',t'';t') =
\frac{1}{V}
\int_V dr'
\frac{\bar{I}}{I_0} [ 1 - \frac{ Vol(ER_1 \cap
IR_2) }{Vol( IR_2)} ]
$$
$$
P_2(r'',t'') =
\int_0^t dt'
\frac{1}{V}
\int_V dr'
\frac{\bar{I}}{I_0} [ 1 - \frac{ Vol(ER_1 \cap
IR_2) }{Vol( IR_2)} ]
$$

Since the probability $P_1(r',t')$ does not depend on $r'$
the integral over $r'$ is not more than a formal object.

It is more convenient to consider the relative deviation of
$P_2$ from the base value $\frac{\bar{I}}{I_0}$
$$
\Delta P_2 (r'',t'';r',t') =
  \frac{ Vol(ER_1 \cap
IR_2) }{Vol( IR_2)}
$$
$$
\Delta P_2(r'',t'';t') =
\frac{1}{V}
\int_V dr'
  \frac{ Vol(ER_1 \cap
IR_2) }{Vol( IR_2)}
$$
$$
\Delta P_2(r'',t'') =
\int_0^t dt'
\frac{1}{V}
\int_V dr'
  \frac{ Vol(ER_1 \cap
IR_2) }{Vol( IR_2)}
$$

The analogous constructions can be made for the three ER
overlapping.

The relative
deviation in probability $\delta
P_3(r''',t''';r'',t'';r',t')$ connected with simultaneous overlapping
of three volumes is given by
$$
\delta P_3(r''',t''';r'',t'';r',t') =
\frac{Vol(ER_1 \cap ER_2 \cap  IR_3)}
{Vol(IR_3)} P_2(r'',t'';r',t')
$$
Here
 there is the deviation in probability to find the
third particle born at $r''',t'''$ when the first particle
was born at $r',t'$ and the second particle was born at
$r'',t''$.

The relative
deviation in probability $\delta
P_3(r''',t''';r'',t'';r',t')$ simply to
find three particles  born in positions $r''',t''';r'',t'';r',t'$
 is given by
$$
\delta P_3(r''',t''';r'',t'';r',t') =
\frac{Vol((ER_1 \cup ER_2) \cap  IR_3)}
{Vol(IR_3)} P_2(r'',t'';r',t')
$$

The knowledge of $P_1, P_2, P3$ solves the problem to find
the rate of nucleation is one-particle, two-particle and
three-particle approximation.

Certainly,
$
ER(t',r')
$
as a function of $t'-t$ instead of $t$ coincides with
$IR(t',r')$

\section{Volumes of overlapping}

In two particle approximation the volume of overlapping
$V_2$ can
be calculated in a very simple way
$$
V_2(r'.t') = \int dt'' \int dr'' Vol(ER_1 \cap ER_2)
P_2(r'',t'';r',t')
$$
Here the argument $r'$ is not important because the system
is uniform.

The partial effective volume in the two-particles
approximation looks like
$$
V_{eff} = Vol(ER_1) - \frac{1}{2} V_2(r',t')
$$

In the three-particles approximation the volume of
overlapping of three volumes is given by
$$
V_{123} = \int dt'' \int dr'' \int t''' \int dr'''
Vol(ER_1 \cap ER_2 \cap ER_3  P_2(r'',t''; r'.t')
P_3 (r'''.t''';r''.t'';r'.t')
$$
It is necessary also to introduce the volumes of
overlapping of ER of two particles in the presence of the
third one. The indication of a particle which does not have
ER in the two-particle overlapping stands in brackets. Then
$$
V_{12(3)} = \int dt'' \int dr'' \int dt''' \int dr'''
Vol(ER_1 \cap ER_2) P_2(r''. t'' ; r',t')
P_3(r''',t''';r'',t'';r',t')
$$

The volume of two-particles overlapping without the volume
of the three-particles overlapping is
approximately given by
$$
\hat{V}_{12(3)} = \int dt'' \int dr'' \int dt''' \int dr'''
Vol((ER_1 \cap ER_2) - (ER_1 \cap ER_2 \cap ER_3))
$$
$$
 P_2(r''. t'' ; r',t')
P_3(r''',t''';r'',t'';r',t')
$$

Analogous value is given by
$$
\hat{V}_{1(2)3} = \int dt'' \int dr'' \int dt''' \int dr'''
Vol((ER_1 \cap ER_3) - (ER_1 \cap ER_2 \cap ER_3))
$$
$$
 P_2(r''. t'' ; r',t')
P_3(r''',t''';r'',t'';r',t')
$$

The effective volume in three particles approximation is
given by
$$
V_{eff} = Vol(ER_1) - V_{12(3)} / 2 -  V_{1(2)3} / 2 - 2
V_{123} / 3
$$

One has to stress that all  volumes indicated as $Vol$ can
be elementary calculated.

\pagebreak

\end{document}